\documentclass[aps,pre,twocolumn,groupedaddress]{revtex4-1}

\usepackage{amsfonts,amssymb,amsmath} 
\usepackage{color} 
\usepackage{graphicx} 
\usepackage{epstopdf,mathrsfs} 
\usepackage[colorlinks,linktocpage,linkcolor=blue]{hyperref}

\begin{document}


\title{Strong anomalous diffusion in two-state process with L\'{e}vy walk and Brownian motion}
\author{Xudong Wang}
\author{Yao Chen}
\author{Weihua Deng}

\affiliation{School of Mathematics and Statistics, Gansu Key Laboratory
of Applied Mathematics and Complex Systems, Lanzhou University, Lanzhou 730000,
P.R. China}



\begin{abstract}
Strong anomalous diffusion phenomena are often observed in complex physical and biological systems, which are characterized by the nonlinear spectrum of exponents $q\nu(q)$ by measuring the absolute $q$-th moment $\langle |x|^q\rangle$. This paper investigates the strong anomalous diffusion behavior of a two-state process with L\'{e}vy walk and Brownian motion, which usually serves as an intermittent search process. The sojourn times in L\'{e}vy walk and Brownian phases are taken as power law distributions with exponents $\alpha_+$ and $\alpha_-$, respectively. Detailed scaling analyses are performed for the coexistence of three kinds of scalings in this system.
Different from the pure L\'{e}vy walk, the phenomenon of strong anomalous diffusion can be observed for this two-state process even when the distribution exponent of L\'{e}vy walk phase satisfies $\alpha_+<1$, provided that $\alpha_-<\alpha_+$. When $\alpha_+<2$, the probability density function (PDF) in the central part becomes a combination of stretched L\'{e}vy distribution and Gaussian distribution due to the long sojourn time in Brownian phase, while the PDF in the tail part (in the ballistic scaling) is still dominated by the infinite density of L\'{e}vy walk.

\end{abstract}


\maketitle

\section{Introduction}
In the recent decades, it is widely recognized that anomalous diffusion is a very general phenomenon in the natural world,
which is characterized by the nonlinear evolution of mean squared displacement with respect to time, i.e., $\langle x^2(t)\rangle \propto t^\beta$ with $\beta\neq1$ \cite{HausKehr:1987,Bouchaud:1992,MetzlerKlafter:2000}. The common examples are $\beta<1$ for subdiffusive continuous-time random walk (CTRW) with divergent first moment of waiting time \cite{BurovJeonMetzlerBarkai:2011,HeBurovMetzlerBarkai:2008} and $\beta>1$ for L\'{e}vy flight with divergent second moment of jump length \cite{ShlesingerZaslavskyFrisch:1995,VahabiSchulzShokriMetzler:2013}. The common feature of the two typical anomalous diffusive processes is their single mode of the motions. However, a particle moving in a complex or even seemingly simple structures might present simultaneous modes \cite{Pikovsky:1991,CastiglioneMazzinoGinanneschiVulpiani:1999}, such as the tracing particle under the effect of a flow acting in the phase space of chaotic Hamiltonian systems \cite{GeiselZacherlRadons:1987,KlafterZumofen:1994}. Such a system is not easy to be analyzed since it exhibits at least two modes of the motion. The common tool to analyze it is the spectrum of exponents $q\nu(q)$ \cite{CastiglioneMazzinoGinanneschiVulpiani:1999} by measuring the absolute $q$-th moment ($q>0$) of the displacement of the particles
\begin{equation}\label{Def-SAD}
  \langle |x(t)|^q\rangle \propto t^{q\nu(q)}.
\end{equation}
For the motions with single mode, $\nu(q)$ is a constant being independent of $q$, such as $\nu(q)\equiv1/2$ for Brownian motion. Otherwise, one can find a nonlinear function $\nu(q)$ of $q$ for the motions with multiple modes; this phenomenon is named as strong anomalous diffusion \cite{CastiglioneMazzinoGinanneschiVulpiani:1999}.

There have been vast systems exhibiting strong anomalous diffusion, such as the nonlinear dynamical systems \cite{CastiglioneMazzinoGinanneschiVulpiani:1999,ArtusoCristadoro:2003,ArmsteadHuntOtt:2003,SandersLarralde:2006,CourbageEdelmanFathiZaslavsky:2008},
the annealed or quenched L\'{e}vy walk \cite{AndersenCastiglioneMazzinoVulpiani:2000,GodrecheLuck:2001,SchmiedebergZaburdaevStark:2009,BurioniCaniparoliVezzani:2010,BernaboBurioniLepriVezzani:2014}, sand pile models \cite{CarrerasLynchNewmanZaslavsky:1999,NewmanSanchezCarrerasFerenbaugh:2002,YadavRamaswamyDhar:2012}, the active transport of polymeric particles in living cells \cite{GalWeihs:2010}, and the spreading of cold atoms in optical lattices \cite{KesslerBarkai:2010,KesslerBarkai:2012,DechantLutz:2012}. The mechanisms of the strong anomalous diffusion for L\'{e}vy walk are studied in detail in Refs. \cite{RebenshtokDenisovHanggiBarkai:2014,RebenshtokDenisovHanggiBarkai:2014-2,RebenshtokDenisovHanggiBarkai:2016}, where the probability density function (PDF) consists of two kinds of distributions---L\'{e}vy distribution in the central part and infinite density in the tail part. The infinite density is non-normalizable, the concept of which was thoroughly investigated as mathematical issues \cite{Aaronson:1997,ThalerZweimuller:2006}, and has been successfully applied to physics; for L\'{e}vy walk, it aims at characterizing the ballistic scaling $(x\sim t)$, which is complementary to the L\'{e}vy scaling in the central part of L\'{e}vy walk.
In contrast, the propagators of subdiffusive CTRW and L\'{e}vy flight only have a single mode, being the stretched Gaussian asymptotics and L\'{e}vy distribution \cite{MetzlerKlafter:2000}, respectively. Compared with L\'{e}vy flight with divergent mean squared displacement, the infinite density characterizes the strong correlation between long jump and long rest in L\'{e}vy walk, resuting in a finite mean squared displacement. In addition, the infinite density can be used to study the rare fluctuations of occupation time statistics in ergodic CTRW \cite{SchulzBarkai:2015} and renewal theory \cite{WangSchulzDengBarkai:2018}.
It is also discussed together with infinite-ergodic theory, for example, the Brownian motion in a logarithmic potential \cite{AghionKesslerBarkai:2019} and the Langevin system with multiplicative noise \cite{LeibovichBarkai:2019,WangDengChen:2019}.

In this paper, we are considering a two-state process alternating between L\'{e}vy walk and Brownian motion, which serves as an intermittent search process \cite{BenichouCoppeyMoreauSuetVoituriez:2005,BenichouLoverdoMoreauVoituriez:2011,LomholtKorenMetzlerKlafter:2008}. The searcher displays a slow active motion in the Brownian phase, during which the hidden target can be detected. While in L\'{e}vy walk phase, the searcher aims to relocate into some unvisited region to reduce oversampling. 
This kind of intermittent search process has wide applications in physical or biological problems \cite{Bell:1991,SongMoonJeonPark:2018,CoppeyBenichouVoituriezMoreau:2004}. The theoretical analyses of the anomalous and nonergodic behavior of this intermittent search process have been investigated in Ref. \cite{WangChenDeng:2019-2}.
Here, we turn our attention to the strong anomalous diffusion behavior of such a two-state process process.
Since the case of pure L\'{e}vy walk has been fully studied in Refs. \cite{RebenshtokDenisovHanggiBarkai:2014,RebenshtokDenisovHanggiBarkai:2014-2,RebenshtokDenisovHanggiBarkai:2016}, the main objective of this paper is to try to discover new phenomena after introducing the Brownian phase in L\'{e}vy walk and to uncover the intrinsic mechanism by clear theoretical analysis.
Intuitively, one can expect that at least three modes coexist in this two-state process. It is true, and furthermore, the newly appeared mode in Brownian phase could bring in many interesting phenomena.
The pure L\'{e}vy walk shows the strong anomalous diffusion phenomenon only in the case of power law exponent $\alpha>1$. Now, the two-state process could exhibit the strong anomalous diffusion even for $\alpha<1$ if the sojourn time in Brownian phase is longer than the one of L\'{e}vy walk phase. Although the particle in Brownian phase could move an arbitrary long distance, the infinite density which characterizes the ballistic scale of L\'{e}vy walk phase with a finite velocity still plays a leading role compared with the Gaussian distribution resulting from Brownian phase.
In particular, another observation different from the pure L\'{e}vy walk is an accumulation effect found at the end of infinite density $(x=\pm v_0t)$.

This paper is organized as follows. In Sec. \ref{two}, we first introduce the two-state process with different power law exponents ($\alpha_+$ and $\alpha_-$) of sojourn time in L\'{e}vy walk and Brownian phases, respectively. Then we derive the corresponding propagator $p_\pm(x,t)$ in two phases in Sec. \ref{three}. The detailed scaling analyses for the cases of $0<\alpha_-<\alpha_+<1$ and $0<\alpha_-<1<\alpha_+$ are presented in Secs. \ref{four} and \ref{five}, respectively. Then in Sec. \ref{six}, we show how these different scaling regimes are complementary and their consistency in the intermediate region. The ensemble-averaged absolute fractional-order moments of the displacement are given in Sec. \ref{seven}. A summary of the key results is provided in Sec. \ref{eight}. In the appendices, some mathematical details are collected.

\section{Model}\label{two}
We consider the process with its motion alternating between two different states---standard L\'{e}vy walk and Brownian motion. For standard L\'{e}vy walk, the particle moves with constant velocity $v_0$ and then changes its direction at a random time. The running times
of each unidirectional flights are independent and drawn from the same distribution. While for Brownian motion, the particle undergoes normal diffusion with diffusivity $D$. Now, we assume that the sojourn time distributions of the two-state process switching between L\'{e}vy walk and Brownian phase are $\psi_+(t)$ and $\psi_-(t)$, respectively. The subscripts `$+$' and `$-$' are introduced to represent the L\'{e}vy walk and Brownian phase, respectively.

This process can be explicitly described by means of the velocity process $v(t)$ which also consists of two states: $v_+(t)$ for L\'{e}vy walk and $v_-(t)$ for Brownian motion. The PDF of $v_+(t)$ is $\delta(|v|-v_0)/2$, while $v_-(t)=\sqrt{2D}\xi(t)$ with $\xi(t)$ being a Gaussian white noise satisfying $\langle\xi(t)\rangle=0$ and $\langle\xi(t_1)\xi(t_2)\rangle=\delta(t_1-t_2)$.

Let the sojourn time distributions in two states be a power law form with exponents $\alpha_\pm$, i.e.,
\begin{equation}
  \psi_\pm(t)\simeq \frac{a_\pm}{|\Gamma(-\alpha_\pm)|t^{1+\alpha_\pm}}
\end{equation}
for large $t$, where $a_\pm$ are scale factors and $\Gamma(\cdot)$ is the Gamma function. The exponents $\alpha_\pm\in(0,2)$ in two states can be the same or different. As usual, we define the Laplace transform $\psi_\pm(s):=\int_0^\infty dt e^{-st}\psi_\pm(t)$
and obtain the asymptotic behavior of $\psi(s)$ for small $s$ as \cite{MiyaguchiAkimotoYamamoto:2016}
\begin{equation}
  \begin{split}
    &\psi_\pm(s)\simeq 1-a_\pm s^{\alpha_\pm},   ~~~~~~~~~~~~~~  \alpha_\pm\in(0,1),  \\
    &\psi_\pm(s)\simeq 1-\mu_\pm s + a_\pm s^{\alpha_\pm},   ~~~~~  \alpha_\pm\in(1,2).
  \end{split}
\end{equation}
For the case of $\alpha_\pm\in(1,2)$, the mean sojourn time, denoted as $\mu_\pm$ for two states, is finite. In particular, the term $s^{\alpha_\pm}$ for $\alpha_\pm\in(1,2)$ is saved to characterize the rare fluctuations of L\'{e}vy walk, i.e., the information in its tail part.
The survival probability of finding the sojourn time in state `$\pm$' exceeding $t$ is defined as $\Psi_\pm(t)=\int_t^\infty dt' \psi_\pm(t')$ with its Laplace transform
\begin{equation}
  \Psi_\pm(s)=\frac{1-\psi_\pm(s)}{s}.
\end{equation}
It is well-known that the dynamical behaviors of standard L\'{e}vy walk vary significantly for different regimes of power law exponents, which naturally motivates us to study the properties of this two-state process with different values of $\alpha_\pm\in(0,2)$.

\section{Propagator of two-state process}\label{three}
The propagator $p(x,t)$ represents the PDF of finding the particle at position $x$ at time $t$, supposing that the particles are initialized at the origin. For this two-state process, we use  $p_\pm(x,t)$ to denote the joint PDF of finding the particle at position $x$ and state `$\pm$' at time $t$. They are associated with the propagator as $p(x,t)=p_+(x, t)+p_-(x,t)$.
Besides, the notation $G_\pm(x,t)$ denotes the conditional PDF of making a displacement $x$ for a complete step in state `$\pm$' within sojourn time $t$, defined as \cite{WangChenDeng:2019-2}
\begin{equation}\label{Gpm}
  \begin{split}
    G_+(x,t)&=\delta(|x|-v_0t)/2,  \\
    G_-(x,t)&=\frac{1}{\sqrt{4\pi Dt}}\exp\left\{-\frac{x^2}{4Dt}\right\},
  \end{split}
\end{equation}
respectively.
Based on these notations and the method of master equations for CTRWs, the integral equations for $p_\pm(x,t)$ can be built as \cite{WangChenDeng:2019-2}
\begin{equation}\label{transport1}
\begin{split}
  \gamma_\pm(x,t)
  &= \int_0^tdt'\int_{-\infty}^{\infty}dx'\psi_\mp(t')G_\mp(x',t')\gamma_\mp(x-x',t-t')  \\
 &~~~+ p^0_\mp\psi_\mp(t)G_\mp(x,t)
\end{split}
\end{equation}
and
\begin{equation}\label{transport2}
\begin{split}
    p_\pm(x,t)
    &=\int_0^t dt'\int_{-\infty}^\infty dx' \Psi_\pm(t')G_\pm(x',t')\gamma_\pm(x-x',t-t') \\
    &~~~+p^0_\pm \Psi_\pm(t)G_\pm(x,t),
\end{split}
\end{equation}
where the flux of particles $\gamma_\pm(x,t)$ defines how many particles leave the position $x$ and change from state `$\mp$' to state `$\pm$' per unit time, and we have taken the initial condition as $p_\pm(x,t=0)=p^0_\pm\delta(x)$ with the constant $p^0_\pm$ being the initial fraction of two states.
By using the techniques of Laplace and Fourier transform
\begin{equation}
  p_\pm(k,s)=\int_0^\infty dt\int_{-\infty}^\infty dx e^{-st}e^{ikx}p_\pm(x,t)
\end{equation}
and performing the transforms on Eqs. \eqref{transport1} and \eqref{transport2}, there are
\begin{equation}\label{transport3}
\begin{split}
\gamma_{\pm}(k,s)&=p_\mp^0\phi_\mp(k,s)+\phi_\mp(k,s)\gamma_{\mp}(k,s),\\
p_{\pm}(k,s)&=p_\pm^0\Phi_{\pm}(k,s)+\Phi_{\pm}(k,s)\gamma_{\pm}(k,s),
\end{split}
\end{equation}
where
\begin{equation}
  \begin{split}
    \phi_+(k,s)&=[\psi_+(s+iv_0k)+\psi_+(s-iv_0k)]/2,  \\[2pt]
    \phi_-(k,s)&=\psi_-(s+Dk^2),   \\[2pt]
    \Phi_+(k,s)&=[\Psi_+(s+iv_0k)+\Psi_+(s-iv_0k)]/2,   \\[2pt]
    \Phi_-(k,s)&=\Psi_-(s+Dk^2).
  \end{split}
\end{equation}
Solving Eq. \eqref{transport3} yields
\begin{equation}\label{pks+-}
\begin{split}
p_{\pm}(k,s) \simeq \frac{\Phi_\pm(k,s)}{1-\phi_+(k,s)\phi_-(k,s)}.
\end{split}
\end{equation}
Based on Eq. \eqref{pks+-}, the explicit expression of propagator $p(k,s)=p_+(k,s)+p_-(k,s)$ can be obtained.
It can be found that the main ingredients of $p(k,s)$ in Eq. \eqref{pks+-} are the sojourn time distributions $\psi_\pm(t)$.
So the further analyses on $p(k,s)$ will be developed for the specific $\psi_\pm(t)$ with fixed $\alpha_\pm$.

Since $\alpha_\pm$ are both in the range $(0,2)$, it can be divided into almost six situations for different values of $\alpha_\pm$: $0<\alpha_-<\alpha_+<1$, $0<\alpha_+<\alpha_-<1$, $0<\alpha_-=\alpha_+<1$, $0<\alpha_+<1<\alpha_-<2$, $0<\alpha_-<1<\alpha_+<2$, and $1<\alpha_\pm<2$. The anomalous and nonergodic behaviors for these six situations have been demonstrated in Ref. \cite{WangChenDeng:2019-2}. The main result therein is that the state with smaller power exponent will dominate the whole process in a power law rate as $t\rightarrow\infty$. In contrast to that, the aim of this paper is to investigate the complementary PDFs and the strong anomalous diffusion behavior of this two-state process.
Comparing with the thoroughly investigated strong anomalous diffusion behavior of standard L\'{e}vy walk, a larger exponent $\alpha_-$ in our two-state process makes no difference on the diffusion behavior.
 Therefore, we only focus on the cases of $0<\alpha_-<\min(\alpha_+,1)$ in this paper, while another four cases will present the same results as pure L\'{e}vy walk.

\section{Scaling analyses for $0<\alpha_-<\alpha_+<1$}\label{four}
In this case, it holds that $\psi_\pm(s)\simeq1-a_\pm s^{\alpha_\pm}$. Substituting it into Eq. \eqref{pks+-}, we obtain the asymptotic form as $(s,k\rightarrow0)$:
\begin{widetext}
\begin{equation}\label{C1-pks}
  p(k,s)\simeq\frac{a_-(s+Dk^2)^{\alpha_--1}+\frac{a_+}{2}[(s+ikv_0)^{\alpha_+-1}+(s-ikv_0)^{\alpha_+-1}]}{a_-(s+Dk^2)^{\alpha_-}+\frac{a_+}{2}[(s+ikv_0)^{\alpha_+}+(s-ikv_0)^{\alpha_+}]}.
\end{equation}
\end{widetext}
The normalization of the propagator $p(x,t)$ can be verified by taking $k=0$ in Eq. \eqref{C1-pks}, which yields $p(0,s)\simeq 1/s$. The direct inverse Fourier-Laplace transform of Eq. \eqref{C1-pks} is infeasible, which implies extra efforts are needed to deal with Eq. \eqref{C1-pks}. Actually, the information contained in the asymptotic form of $p(k,s)$ could be extracted through some appropriate scaling analyses. By carefully looking at the denominator in Eq. \eqref{C1-pks}, three kinds of scaling ($s\sim k$, $s\sim |k|^2$, and $s\sim |k|^{\alpha_+/\alpha_-}$) can be observed. We will first consider the scaling $s\sim k$, since it characterizes the ballistic scaling of L\'{e}vy walk due to its unidirectional flight at each sojourn in this phase.

The outmost distance that particles can arrive at is $\pm v_0t$ in L\'{e}vy walk phase, linear with time, which truncates the PDF $p(x,t)$ at $\pm v_0t$. Although the particle in Brownian phase might go farther than the distance $\pm v_0t$, the corresponding distribution decays exponentially when $x\gg \sqrt{t}$ as the propagator $G_-(x,t)$ shows in Eq. \eqref{Gpm}. So we omit the contributions of Brownian phase in the scaling $s\sim k$.
The explicit tail information is described by the ballistic scaling $s\sim k$ in Eq. \eqref{C1-pks}, corresponding to $x\sim t$ in space-time domain. In contrast to $s\sim k$, another two kinds of scalings aim at characterizing the central part of the graph of the PDF $p(x,t)$.

\subsection{Infinite density of rare fluctuations}
To consider the scaling $s\sim k$, we let $s,k\rightarrow0$ and $s/k$ be fixed. Then Eq. \eqref{C1-pks} can be rewritten as
\begin{equation}\label{C1-pks2}
\begin{split}
    &p(k,s)\simeq  \frac{1}{s} \\
    &~~\times\frac{1+\frac{a_+}{2a_-}s^{\alpha_+-\alpha_-}
    [(1+\frac{ikv_0}{s})^{\alpha_+-1}+(1-\frac{ikv_0}{s})^{\alpha_+-1}]}
    {1+\frac{a_+}{2a_-}s^{\alpha_+-\alpha_-}[(1+\frac{ikv_0}{s})^{\alpha_+}+(1-\frac{ikv_0}{s})^{\alpha_+}]}
\end{split}
\end{equation}
after neglecting the higher order term $k^2$. The two terms in Eq. \eqref{C1-pks2} containing $s^{\alpha_+-\alpha_-}$ tend to zero since $\alpha_+>\alpha_-$. Thus, we further have the asymptotic form as
\begin{equation}\label{C1-pks3}
\begin{split}
    p(k,s)&\simeq \frac{1}{s} +\frac{a_+}{2a_-}s^{\alpha_+-\alpha_--1} \\
    &~~~\times\left[\left(1+\frac{ikv_0}{s}\right)^{\alpha_+-1}+\left(1-\frac{ikv_0}{s}\right)^{\alpha_+-1}\right. \\
     &~~~~~~~~ \left. -\left(1+\frac{ikv_0}{s}\right)^{\alpha_+}-\left(1-\frac{ikv_0}{s}\right)^{\alpha_+}\right]  \\
      &= \frac{1}{s}+\frac{a_+}{2a_-} [R_\alpha(k,s)+R_\alpha(-k,s)],
\end{split}
\end{equation}
where for convenience we use the notation:
\begin{equation}\label{C1-Rks}
  \begin{split}
    &R_\alpha(k,s) \\
    &:= s^{\alpha_+-\alpha_--1}\left[\left(1+\frac{ikv_0}{s}\right)^{\alpha_+-1}-\left(1+\frac{ikv_0}{s}\right)^{\alpha_+}\right]  \\
    &~= \frac{-ikv_0}{s^{\alpha_-+1}}(s+ikv_0)^{\alpha_+-1}.
  \end{split}
\end{equation}
The leading term in Eq. \eqref{C1-pks3} is $1/s$, the inverse Fourier-Laplace transform of which is $\delta(x)$. It contributes to a normalized PDF in this scaling while the latter two terms provide the information on the tail of the PDF $p(x,t)$. We consider the ballistic scaling $s\sim k$ here, which compresses all the information in the central part into the origin and thus yields the normalized term $\delta(x)$. Since we are focusing on the information in the tail $|x|>0$, we omit the term $\delta(x)$ and pay attention to the inverse of $R_\alpha(\pm k,s)$ in Eq. \eqref{C1-pks3}. With some technical calculations in Appendix \ref{App1}, the inversion of $R_\alpha(\pm k,s)$ is obtained and there is
\begin{equation}\label{pxt-tail1}
  \begin{split}
    p(x,t) \simeq \frac{a_+}{2a_-v_0}\frac{t^{\alpha_--\alpha_+-1}}{\Gamma(\alpha_-+1)\Gamma(1-\alpha_+)} \mathcal{I}\left(\frac{|x|}{v_0t}\right),
  \end{split}
\end{equation}
where
\begin{equation}\label{Iz}
  \mathcal{I}(z) = \textbf{1}_{(0<z\leq1)} z^{-\alpha_+-1}(1-z)^{\alpha_--1}\left[\alpha_++(\alpha_--\alpha_+)z \right].
\end{equation}

There is a truncation at $z=1$ in the expression of $\mathcal{I}(z)$, which implies $|x|\leq v_0t$, consistent to the previous analysis that the particle will not go beyond the distance $\pm v_0t$. On the other hand, regarding $z=x/v_0t$ as a new variable, the integral of the auxiliary function $\mathcal{I}(z)$ diverges due to its singularity at the origin $z=0$, which gives it a name---infinite density. Therefore, the infinite density $\mathcal{I}(z)$ is not a real physical PDF. Despite of this, it reveals the long time asymptotic behavior of the propagator $p(x,t)$ through the relationship in Eq. \eqref{pxt-tail1}. When calculating moments, we multiply $|x|^q$ on both sides of Eq. \eqref{pxt-tail1} and integrate with respect to $x$. Then we obtain
\begin{equation}\label{relation-ID}
  \int_{-\infty}^\infty |x|^qp(x,t)dx \propto t^{\alpha_--\alpha_++q}\int_0^{1}z^q \mathcal{I}(z)dz.
\end{equation}
For $q<\alpha_+$, the integral on the right-hand side of Eq. \eqref{relation-ID} diverges. However, the infinite density $\mathcal{I}(z)$ is valid for high order moments with $q>\alpha_+$,  which cures the singularity at $z=0$.
Therefore, the main functions of the infinite density $\mathcal{I}(z)$ is to characterize the tail information of PDF $p(x,t)$ and to calculate the high order moments.

We also observe another interesting phenomenon---an accumulation at $z=1$ due to $\alpha_-<1$ in Eq. \eqref{Iz}. This accumulation even exists for $D=0$ (i.e., the L\'{e}vy walk interrupted by rest \cite{SolomonWeeksSwinney:1993,KlafterZumofen:1994,SongMoonJeonPark:2018}). Therefore, this accumulation is not contributed by the particles in Brownian phase, which vanishes when $\alpha_-=1$. While for $\alpha_-<1$ and big $t$, it can be balanced by the prefactor $t^{\alpha_--\alpha_+-1}$ in Eq. \eqref{pxt-tail1}. Actually, this phenomenon implies that the PDF of pure L\'{e}vy walk is dropped down by the long sojourn time in Brownian phase except for the end point at $z=1$.
The end point of the infinite density is not affected since it results from the particles running in its first step for the whole time. Once the particle renews and turns into the second step in Brownian phase with longer sojourn time, it is less likely for the particle to go back to the L\'{e}vy walk phase again.

\subsection{Dual scaling regimes in the central part}
After obtaining the tail information of $p(x,t)$ by introducing an infinite density $\mathcal{I}(z)$, we turn our attention to the central part of $p(x,t)$ where the scaling relation $s\ll k$ is valid. This scaling helps to simplify the Eq. \eqref{C1-pks} into
\begin{equation}\label{C1-pks4}
\begin{split}
    &p(k,s)\simeq \\
    &~~\frac{a_- (s+Dk^2)^{\alpha_--1}+\frac{a_+}{2}[(ikv_0)^{\alpha_+-1}+(-ikv_0)^{\alpha_+-1}]}{a_- (s+Dk^2)^{\alpha_-}+\frac{a_+}{2}[(ikv_0)^{\alpha_+}+(-ikv_0)^{\alpha_+}]}.
\end{split}
\end{equation}
It can be found that two different scalings coexist in Eq. \eqref{C1-pks4}, i.e., $s\sim|k|^{\alpha_+/\alpha_-}$ and $s\sim|k|^2$. This phenomenon is different from the standard L\'{e}vy walk, where only L\'{e}vy scaling is observed at the central part \cite{RebenshtokDenisovHanggiBarkai:2014-2}. Now the Gaussian shape with scaling $s\sim|k|^2$ cannot be omitted due to the longer sojourn time in Brownian phase.

Therefore, it is necessary to consider the magnitude relation between $\alpha_+$ and $2\alpha_-$ for further analyses. If $\alpha_+<2\alpha_-$, the L\'{e}vy scaling $s\sim|k|^{\alpha_+/\alpha_-}$ dominates the PDF $p(x,t)$. In this case, after omitting $Dk^2$ and the second term in numerator of Eq. \eqref{C1-pks4} due to $|k|^{\alpha_+-1}\ll s^{\alpha_--1}$, we obtain
\begin{equation}\label{C1-pks5}
\begin{split}
    p(k,s)&\simeq
    \frac{a_- s^{\alpha_--1}}{a_- s^{\alpha_-}+ a_+\cos(\pi\alpha_+/2)v_0^{\alpha_+}|k|^{\alpha_+}} \\
    &= \frac{s^{\alpha_--1}}{s^{\alpha_-}+K_\alpha|k|^{\alpha_+}},
\end{split}
\end{equation}
where the generalized diffusion coefficient $K_\alpha=a_+\cos(\pi\alpha_+/2)v_0^{\alpha_+}/a_-$. When $\alpha_-=1$, the inverse of $p(k,s)$ is a normalized symmetric L\'{e}vy stable PDF, which recovers the central part of the PDF of standard L\'{e}vy walk. For $\alpha_-<1$, the PDF is like a L\'{e}vy flight coupled with an inverse subordinator. The displacement in Brownian phase can be neglected and thus it acts like a trap event with power law exponent $\alpha_-<1$. The running time in L\'{e}vy walk phase is far less than the one in Brownian phase and thus can be neglected so that the displacement in this phase acts like a jump obeying power law distribution with exponent $\alpha_+$. The corresponding Langevin system can be found in Ref. \cite{Fogedby:1994}.
The PDF $p(x,t)$ in Eq. \eqref{C1-pks5} is a stretched L\'{e}vy distribution, the closed form of which can be expressed by Fox H-function:
\begin{equation*}
  p(x,t)\simeq \frac{1}{\sqrt{\pi}|x|} H^{2,1}_{2,3}
        \left[\left.\frac{|x|^{\alpha_+}}{2^{\alpha_+}K_\alpha t^{\alpha_-}}\right|
        \begin{array}{l}
          (1,1),(1,\alpha_-)  \\ (\frac{1}{2},\frac{\alpha_+}{2}),(1,1),(1,\frac{\alpha_+}{2})
        \end{array}\right].
\end{equation*}
Based on the asymptotic form of Fox H-function \cite{MathaiSaxenaHaubold:2009}, there is
\begin{equation}\label{pxt-cent1}
  p(x,t)\simeq \frac{\tilde{K}_\alpha t^{\alpha_-}}{|x|^{1+\alpha_+}}
\end{equation}
for large $|x|$, where the coefficient $$\tilde{K}_\alpha=\frac{\Gamma(1+\alpha_+)\sin(\pi\alpha_+/2)K_\alpha}{\Gamma(1+\alpha_-)\pi}.$$

On the contrary, if $\alpha_+>2\alpha_-$, the dominant part of $p(k,s)$ is in the scaling $s\sim |k|^2$. Similarly, the second terms in numerator and denominator of Eq. \eqref{C1-pks4} are both higher order than the corresponding first terms. We neglect them and obtain
\begin{equation}\label{C1-pks6}
  p(k,s)\simeq \frac{1}{s+Dk^2},
\end{equation}
displaying the classical behavior of Brownian motion. The inverse Fourier-Laplace transform of Eq. \eqref{C1-pks6} yields the Gaussian shape
\begin{equation}\label{pxt-cent2}
  p(x,t)\simeq \frac{1}{\sqrt{4\pi Dt}}\exp\left(-\frac{x^2}{4Dt}\right)
\end{equation}
in the central part of $p(x,t)$.

Compared with the infinite density $\mathcal{I}(z)$, the different asymptotic forms of $p(x,t)$ on the central part within different scaling regimes are both normalized, since taking $k=0$ both yield $p(0,s)\simeq 1/s$ in Eqs. \eqref{C1-pks5} and \eqref{C1-pks6}. However, the high order (bigger than $\alpha_+$) moments will diverge if we use the asymptotic forms of $p(x,t)$ in the central part, since the large-$x$ behavior in Eq. \eqref{pxt-cent1} is heavy-tailed with exponent $1+\alpha_+$.
On the contrary, the high order moments with Gaussian PDF in Eq. \eqref{pxt-cent2} exponentially decay  and can be neglected compared with the infinite density $\mathcal{I}(z)$ on the tail part.

\section{Scaling analyses for $0<\alpha_-<1<\alpha_+$}\label{five}

In this case, it holds that $\psi_+(s)\simeq1-\mu_+ s +a_+ s^{\alpha_+}$ and $\psi_-(s)\simeq1-a_- s^{\alpha_-}$. Substituting them into Eq. \eqref{pks+-}, we obtain the asymptotic form as $(s,k\rightarrow0)$:
\begin{widetext}
\begin{equation}\label{C2-pks}
  p(k,s)\simeq\frac{\mu_++a_-(s+Dk^2)^{\alpha_--1}-\frac{a_+}{2}[(s+ikv_0)^{\alpha_+-1}+(s-ikv_0)^{\alpha_+-1}]}{\mu_+s+a_-(s+Dk^2)^{\alpha_-}-\frac{a_+}{2}[(s+ikv_0)^{\alpha_+}+(s-ikv_0)^{\alpha_+}]},
\end{equation}
\end{widetext}
which is also normalized. It can be found that the slight difference between Eqs. \eqref{C2-pks} and \eqref{C1-pks} are the terms containing $\mu_+$. Considering $\alpha_-<1$, there is $\mu_+\ll(s+Dk^2)^{\alpha_--1}$. Therefore, $\mu_+$ can be omitted in the denominator and numerator of Eq. \eqref{C2-pks}. Then the asymptotic form in Eq. \eqref{C2-pks} is almost the same as the one in Eq. \eqref{C1-pks}, except for the minuses in front of the last terms in the denominator and numerator.

Although the asymptotic forms are similar in Eqs. \eqref{C1-pks} and \eqref{C2-pks}, the details of scaling analyses are slightly different due to $\alpha_+>1$. More precisely, let us first focus on the scaling $s\sim k$. A result similar to Eq. \eqref{C1-pks3} can be obtained as
\begin{equation}\label{C2-pks3}
\begin{split}
    p(k,s)= \frac{1}{s}-\frac{a_+}{2a_-} [R_\alpha(k,s)+R_\alpha(-k,s)].
\end{split}
\end{equation}
Since $\alpha_+>1$ now, we need to split $R_\alpha(k,s)$ into two parts to perform inverse Fourier-Laplace transforms, that is,
\begin{equation}
  R_\alpha(k,s) = \left[\frac{-ikv_0}{s^{\alpha_-}}+\frac{v_0^2k^2}{s^{\alpha_-+1}}\right](s+ikv_0)^{\alpha_+-2}.
\end{equation}
With similar procedures as Appendix \ref{App1}, we finally get
\begin{equation}\label{pxt-tail2}
  \begin{split}
    p(x,t) \simeq \frac{a_+}{2a_-v_0}\frac{t^{\alpha_--\alpha_+-1}}{\Gamma(\alpha_-+1)|\Gamma(1-\alpha_+)|} \mathcal{I}\left(\frac{|x|}{v_0t}\right),
  \end{split}
\end{equation}
where $\mathcal{I}(z)$ is defined in Eq. \eqref{Iz}. To replace $\Gamma(1-\alpha_+)$ by $|\Gamma(1-\alpha_+)|$ for positivity preserving, one can get Eq. \eqref{pxt-tail2} from Eq.  \eqref{pxt-tail1}, which is the only difference.
Similarly, for the scaling analyses when $s\ll k$, Eqs. \eqref{C1-pks5} and \eqref{C1-pks6} are also valid if we replace $\cos(\pi\alpha_+/2)$ by $|\cos(\pi\alpha_+/2)|$.

\begin{figure*}[tbhp]
\begin{minipage}{0.33\linewidth}
  \centerline{\includegraphics[scale=0.4]{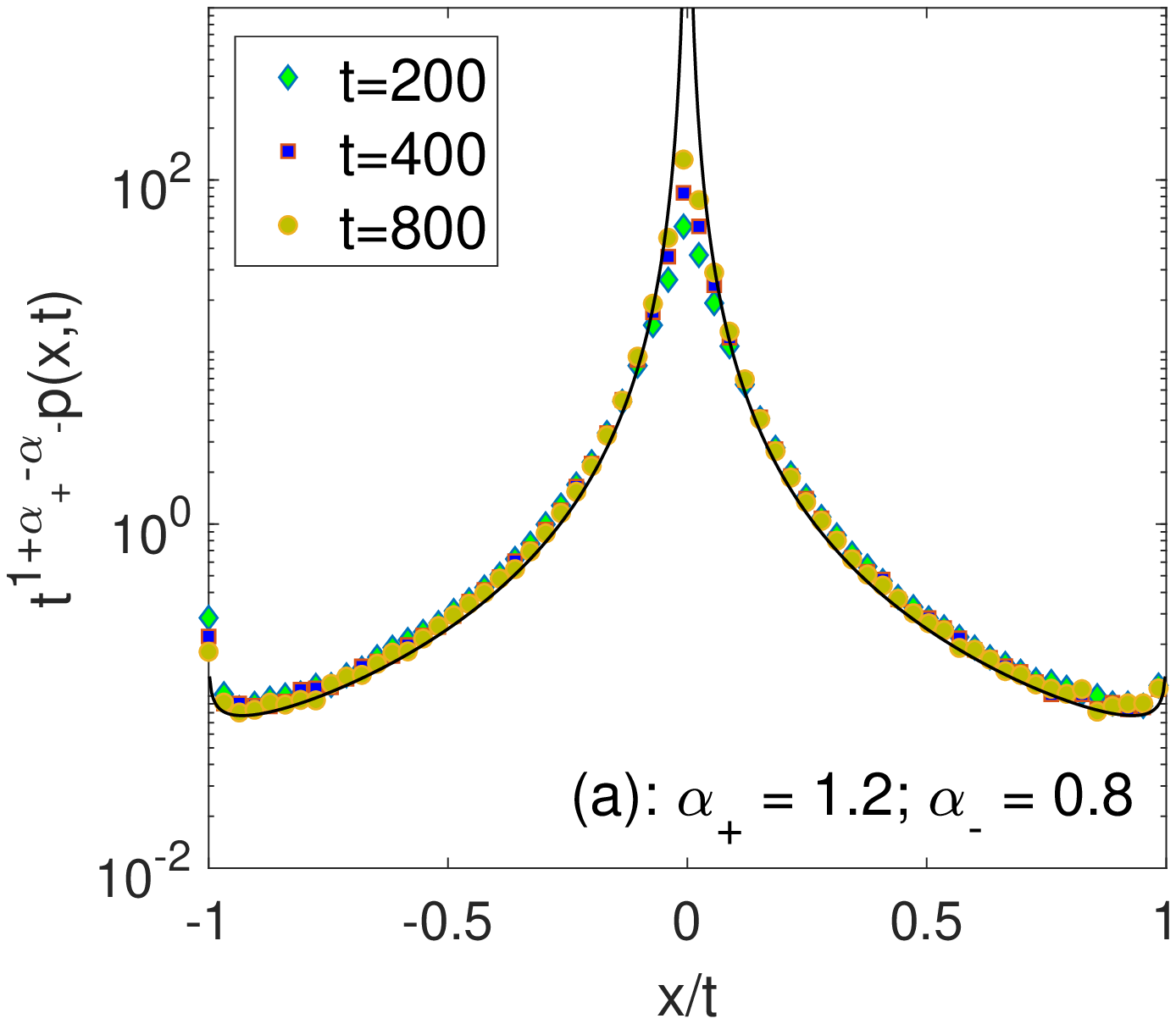}}
\end{minipage}
\hspace{-2mm}
\begin{minipage}{0.33\linewidth}
  \centerline{\includegraphics[scale=0.4]{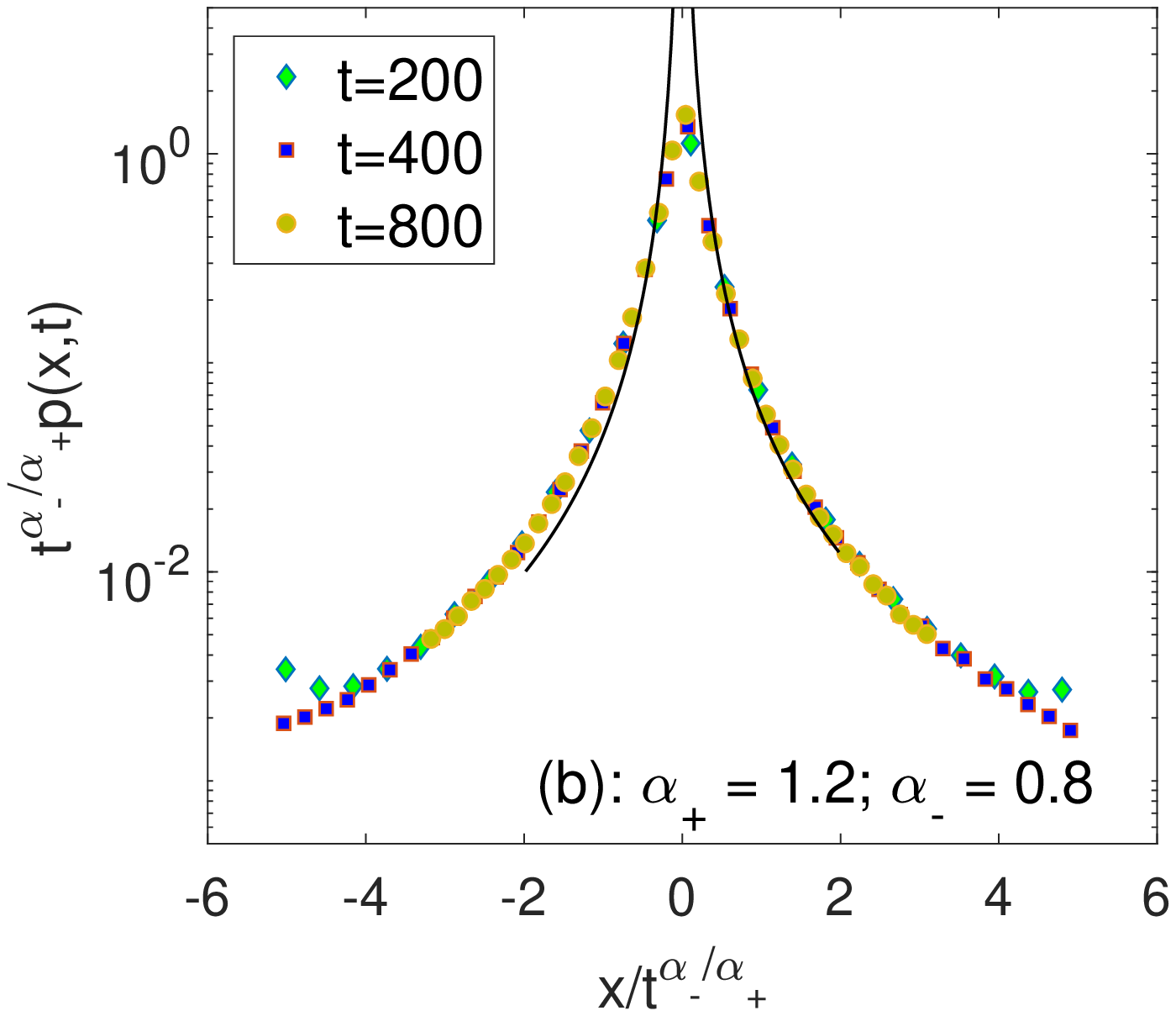}}
\end{minipage}
\hspace{-2mm}
\begin{minipage}{0.33\linewidth}
  \centerline{\includegraphics[scale=0.4]{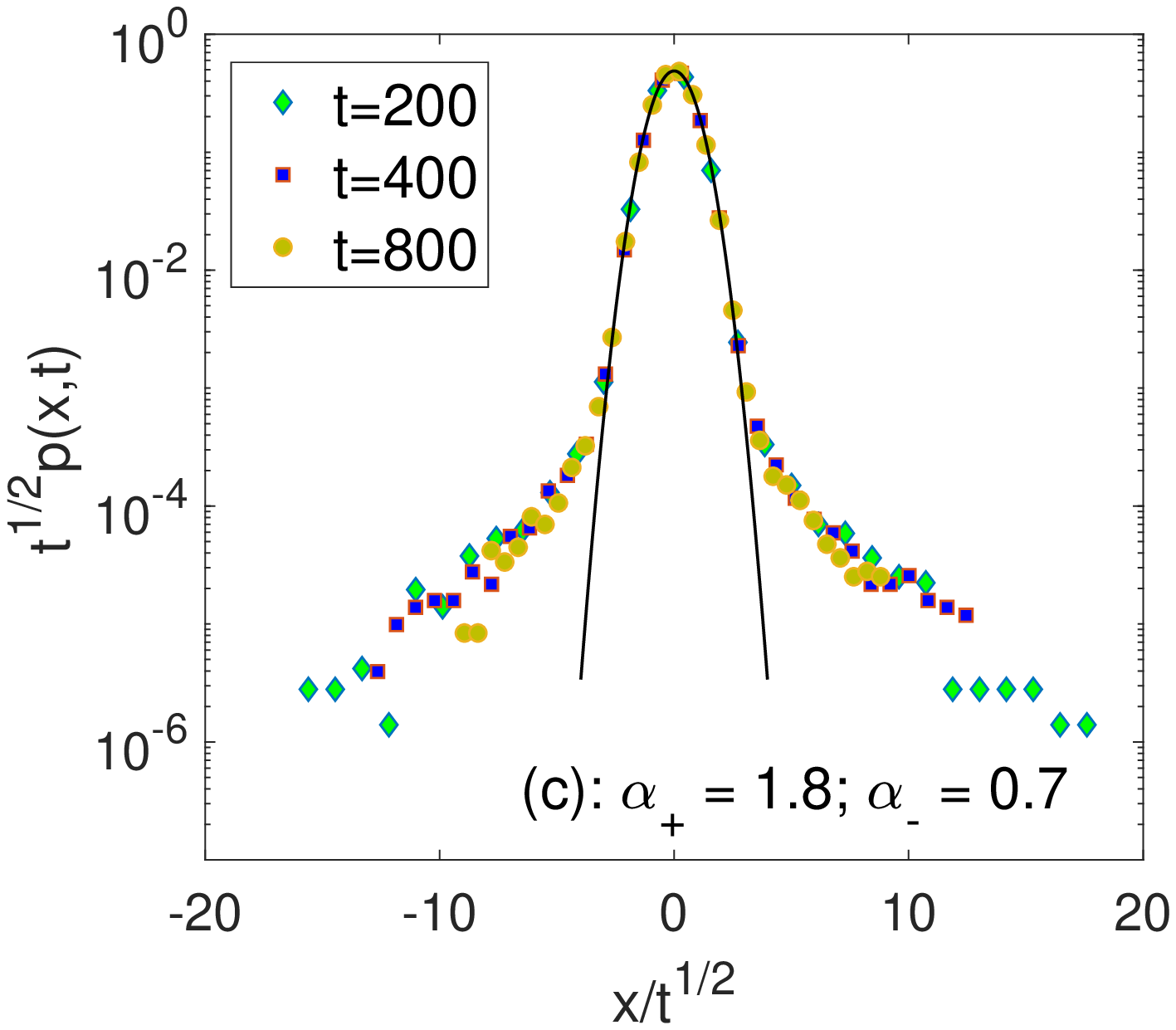}}
\end{minipage}

\vspace{1pt}
\begin{minipage}{0.33\linewidth}
  \centerline{\includegraphics[scale=0.4]{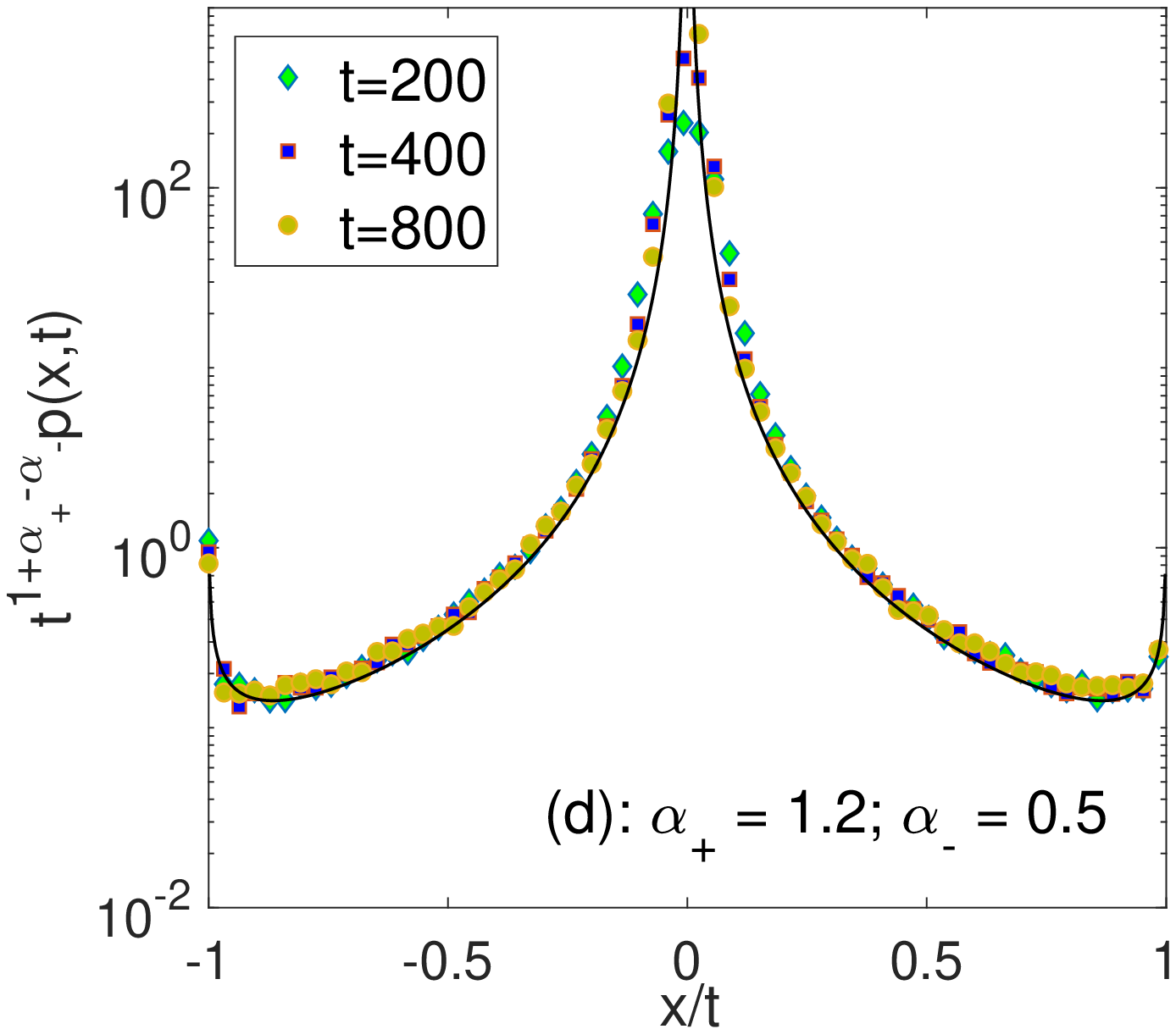}}
\end{minipage}
\hspace{-2mm}
\begin{minipage}{0.33\linewidth}
  \centerline{\includegraphics[scale=0.39]{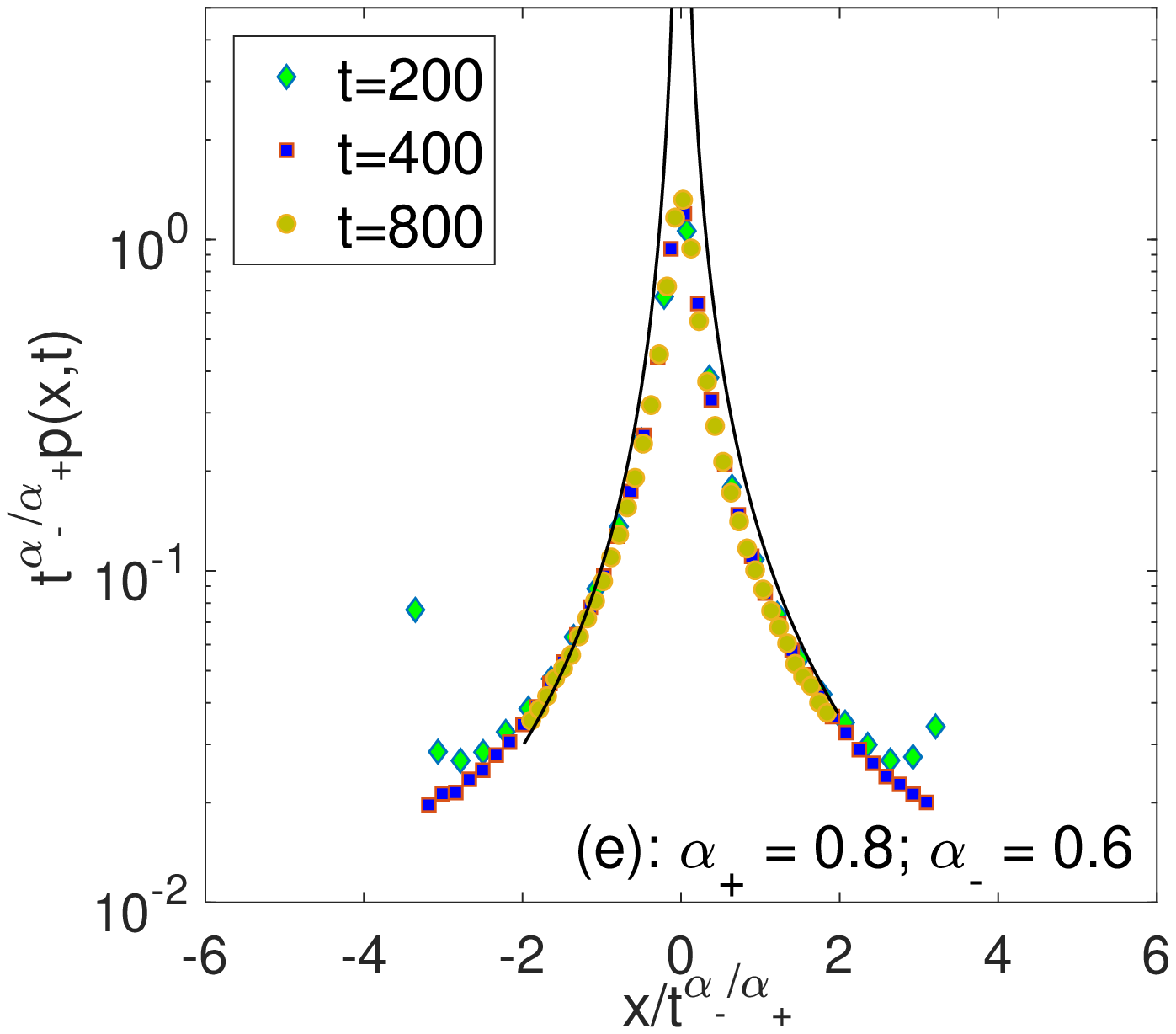}}
\end{minipage}
\hspace{-2mm}
\begin{minipage}{0.33\linewidth}
  \centerline{\includegraphics[scale=0.4]{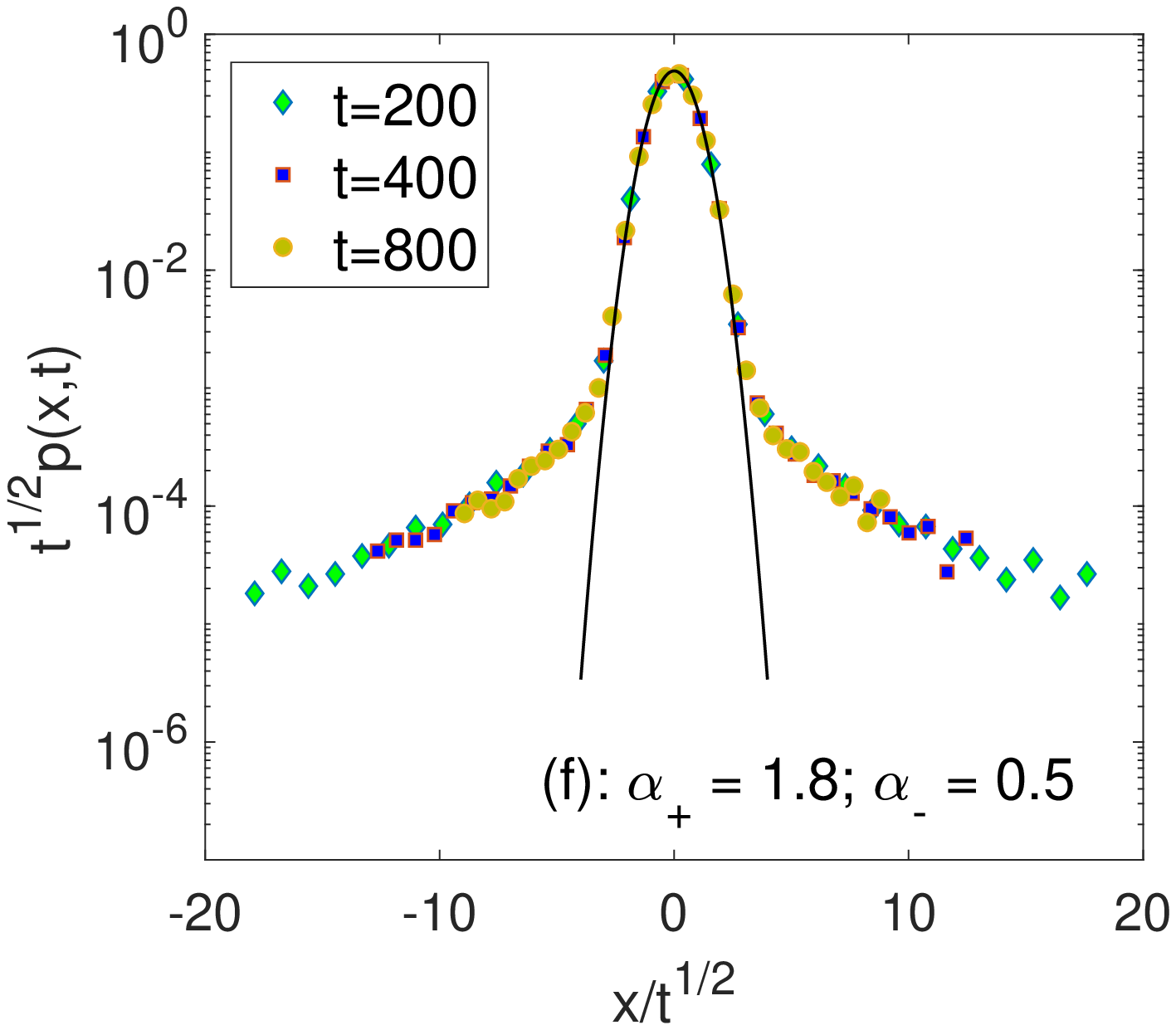}}
\end{minipage}
\caption{Scaled PDF of two-state process. Color symbols represent the simulation results while the black solid lines are the theoretical results with parameters $v_0=1$, $D=0.1$, and $\tau_0=0.1$. All the simulations agree with the theoretical results very well.
In (a) and (d), $\alpha_+=1.2,\alpha_-=0.8$ and $\alpha_+=1.2,\alpha_-=0.5$ are chosen for the case $\alpha_+<2\alpha_-$ and $\alpha_+>2\alpha_-$ to verify the infinity density $g_{\textrm{tail}}(z)$ in Eq. \eqref{comp-tail}, respectively. In (b) and (e), $\alpha_+=1.2,\alpha_-=0.8$ and $\alpha_+=0.8,\alpha_-=0.6$ are taken for the case $\alpha_+<2\alpha_-$ to verify $g_{\textrm{cen1}}(z)$ in Eq. \eqref{solu-cen}. The solid line describes the asymptotic result of the Fox H-function for large $z$, i.e., $g_{\textrm{cen1}}(z)\simeq \tilde{K}_\alpha |z|^{-1-\alpha_+}.$ In (c) and (f), $\alpha_+=1.8,\alpha_-=0.7$ and $\alpha_+=1.8,\alpha_-=0.5$ for the case $\alpha_+>2\alpha_-$ to verify $g_{\textrm{cen2}}(z)$ in Eq. \eqref{solu-cen}.  }\label{fig1}
\end{figure*}

\section{complementarity among different scaling regimes}\label{six}
 For both cases of $0<\alpha_-<\alpha_+<1$ and $0<\alpha_-<1<\alpha_+$, the PDFs $p(x,t)$ are studied in different scaling regimes. The tail part ($x\sim t$) can be well approximated by the infinity density $\mathcal{I}(z)$ as
\begin{equation}\label{solu-tail}
  p(x,t)\simeq t^{\alpha_--\alpha_+-1}g_{\textrm{tail}}\left(\frac{x}{t}\right),
\end{equation}
where the scaling function
\begin{equation}\label{comp-tail}
  g_{\textrm{tail}}(z)=\frac{a_+}{2a_-v_0\Gamma(\alpha_-+1)|\Gamma(1-\alpha_+)|} \mathcal{I}\left(\frac{|z|}{v_0}\right).
\end{equation}
On the other hand, the central part of $p(x,t)$ is well approximated by two densities as
\begin{equation}\label{solu-cen}
  p(x,t)\simeq \left\{
  \begin{array}{ll}
    t^{-\alpha_-/\alpha_+} \,g_{\textrm{cen1}}\left(xt^{-\alpha_-/\alpha_+}\right)
    &~ \alpha_+<2\alpha_-, \\[5pt]
    t^{-1/2} \,g_{\textrm{cen2}}\left(xt^{-1/2}\right)  &~ \alpha_+>2\alpha_-,
  \end{array} \right.
\end{equation}
where the scaling functions
\begin{equation*}
  g_{\textrm{cen1}}(z)=\frac{1}{\sqrt{\pi}|z|} H^{2,1}_{2,3}
        \left[\left.\frac{|z|^{\alpha_+}}{2^{\alpha_+}K_\alpha }\right|
        \begin{array}{l}
          (1,1),(1,\alpha_-)  \\ (\frac{1}{2},\frac{\alpha_+}{2}),(1,1),(1,\frac{\alpha_+}{2})
        \end{array}\right]
\end{equation*}
and
\begin{equation*}
  g_{\textrm{cen2}}(z)=\frac{1}{\sqrt{4\pi D}}\exp\left(-\frac{z^2}{4D}\right).
\end{equation*}

The central part of $p(x,t)$ is with the scaling $x\sim t^\beta$, where $$\beta=\max(\alpha_-/\alpha_+,1/2)<1.$$ Therefore, the intermediate region between central part $t^\beta$ and tail part $t$ is very large as $t\rightarrow\infty$. For convenience, we simplify Eq. \eqref{solu-cen} as
\begin{equation}
  p(x,t)\simeq  t^{-\beta}\,g_{\textrm{cen}}\left(\frac{x}{t^{\beta}}\right),
\end{equation}
where $g_{\textrm{cen}}=g_{\textrm{cen1}}$ when $\alpha_+<2\alpha_-$ and $g_{\textrm{cen}}=g_{\textrm{cen2}}$ when $\alpha_+>2\alpha_-$.
To verify the results of PDF $p(x,t)$ in Eqs. \eqref{solu-tail} and \eqref{solu-cen}, we simulate the PDF with different scalings for several pairs of $\alpha_\pm$. The simulation results are presented in Fig. \ref{fig1}, showing the agreement with theoretical results very well.

A natural expectation on the analyses in different scales is that the different distributions should be consistent in the intermediate region. The intermediate region is described by $x\rightarrow0$ for tail part and $x\rightarrow\infty$ for central part. By taking the corresponding limits in Eqs. \eqref{solu-tail} and \eqref{solu-cen}, respectively, we obtain the same asymptotic form as
\begin{equation}\label{intermediate}
  \begin{split}
    p(x,t)&\simeq t^{\alpha_--\alpha_+-1}g_{\textrm{tail}}\left(\frac{x}{t}\right)  \\
    &\simeq t^{-\alpha_-/\alpha_+} \,g_{\textrm{cen1}}\left(\frac{x}{t^{\alpha_-/\alpha_+}}\right) \\
    &\simeq \frac{c_0t^{\alpha_-}}{|x|^{1+\alpha_+}}
  \end{split}
\end{equation}
for $t^\beta\ll x\ll t$,
where the coefficient $$c_0=\frac{a_+\alpha_+v_0^{\alpha_+}}{2a_-\Gamma(1+\alpha_-)|\Gamma(1-\alpha_+)|}.$$
The different scaling regimes are complementary here, and they together depict the whole graph of PDF $p(x,t)$. Note that we only use the first density in the central part which is power law decay in Eq. \eqref{solu-cen}, since another one decays exponentially and can be omitted.
Apart from the consistence of two distributions in the intermediate region,
the previous discussions of different dominant roles in Eq. \eqref{solu-cen} make sense when calculating low order moments.

\section{Ensemble averages}\label{seven}
Now we pay attention to the absolute moments of all orders for the displacement.
Since the different scaling regimes approximate the different parts of $p(x,t)$, they together yield the entire information on the long time asymptotics, and thus determine the moments of displacement. We introduce an auxiliary function $c(t)$ which satisfies
\begin{equation}
  t^\beta\ll c(t)\ll t
\end{equation}
to divide the central part and tail part. Then we can split the integral into two parts, where different scaling regimes  well approximate $p(x,t)$, that is,
\begin{equation}\label{moments}
  \begin{split}
    \langle |x(t)|^q\rangle &= \int_{|x|\leq c(t)}|x|^qp(x,t)dx+\int_{|x|> c(t)}|x|^qp(x,t)dx \\
    &=\int_{|x|\leq c(t)}|x|^q t^{-\beta}\,g_{\textrm{cen}}\left(\frac{x}{t^{\beta}}\right)dx \\
    &~~~~~+\int_{|x|> c(t)}|x|^q  t^{\alpha_--\alpha_+-1}g_{\textrm{tail}}\left(\frac{x}{t}\right)dx \\
    &=  t^{\beta q} \int_{|z|\leq c(t)/t^\beta}|z|^q g_{\textrm{cen}}(z)dz \\
    &~~~~~+t^{\alpha_--\alpha_++q} \int_{|z|> c(t)/t}|z|^q g_{\textrm{tail}}(z)dz.
  \end{split}
\end{equation}
Therefore, the central and tail parts have different contributions to the absolute $q$-th moments, which are $t^{\beta q}$ and $t^{\alpha_--\alpha_++q}$, respectively, the critical value of which is
\begin{equation}
  q_c=\frac{\alpha_+-\alpha_-}{1-\beta},
\end{equation}
implying the piecewise linear behavior of the spectrum of exponents $q\nu(q)$ in Eq. \eqref{Def-SAD}.
When $q<q_c$, the former one plays a leading role, otherwise the latter one dominates. The two integrals in Eq. \eqref{moments} are both finite by choosing appropriate $c(t)$ for different order $q$. For example, for low order moments with $q<q_c$, choosing $c(t)=c_1t$ with $c_1\ll1$, the two integrals become
\begin{equation}
  \int_{-\infty}^\infty|z|^q g_{\textrm{cen}}(z)dz,~~\int_{|z|> c_1}|z|^q g_{\textrm{tail}}(z)dz < \infty
\end{equation}
as $t\rightarrow\infty$. The singular point $z=0$ of the latter integral is excluded by a small distance $c_1$.
While for high order moments with $q>q_c$, we choose $c(t)=c_2t^\beta$ with $1\ll c_2$. Then the two integrals are
\begin{equation}
  \int_{|z|\leq c_2}|z|^q g_{\textrm{cen}}(z)dz,~~\int_{-\infty}^\infty|z|^q g_{\textrm{tail}}(z)dz < \infty.
\end{equation}
The high order moments for the infinity density $g_{\textrm{tail}}(z)$ will not diverge.

\begin{figure}
  \centering
  \includegraphics[scale=0.55]{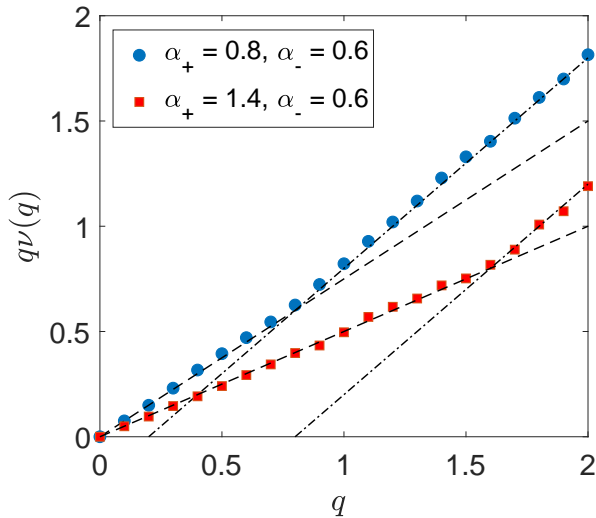}\\
  \caption{The spectrum of exponents $q\nu(q)$ versus $q$. The blue circle-markers and red square-markers represent the simulation results of the sampling with $10^5$ realizations for cases $\alpha_+=0.8,\alpha_-=0.6$ and $\alpha_+=1.4,\alpha_-=0.6$, respectively.
  The dotted lines and dot-dashed lines denote the theoretical results for $q<q_c$ and $q>q_c$ in Eqs. \eqref{EA_theo1} and \eqref{EA_theo2}, respectively.}\label{fig2}
\end{figure}

Considering $\beta=\max(\alpha_-/\alpha_+,1/2)$, the absolute $q$-th moments are given in two different cases.
If $\alpha_+<2\alpha_-$,
\begin{equation}\label{EA_theo1}
\begin{split}
    \langle |x(t)|^q\rangle \simeq
    \left\{
      \begin{array}{ll}
        M^{\prec}_1 \, t^{q\alpha_-/\alpha_+}, & q<\alpha_+, \\
        M^{\succ} \, t^{q+\alpha_--\alpha_+}, & q>\alpha_+.
      \end{array}
    \right.
\end{split}
\end{equation}
If $\alpha_+\ge 2\alpha_-$,
\begin{equation}\label{EA_theo2}
\begin{split}
    \langle |x(t)|^q\rangle \simeq
    \left\{
      \begin{array}{ll}
       M^{\prec}_2 \, t^{q/2}, & q<2(\alpha_+-\alpha_-), \\
       M^{\succ} \, t^{q+\alpha_--\alpha_+}, & q>2(\alpha_+-\alpha_-).
      \end{array}
    \right.
\end{split}
\end{equation}
The results in Eqs. \eqref{EA_theo1} and \eqref{EA_theo2} imply that this system exhibits strong anomalous diffusion behavior with a bilinear spectrum of exponents, which has been verified by simulations in Fig. \ref{fig2}.
The diffusion coefficients $M^{\prec}_1$, $M^{\prec}_2$, and $M^{\succ}$ can be obtained from the derivations in Eq. \eqref{moments} as
\begin{equation}\label{M-coefficients}
  \begin{split}
    M^{\prec}_1&=\int_{-\infty}^\infty|z|^q g_{\textrm{cen1}}(z)dz  \\
    &=\frac{(K_\alpha)^{q/\alpha_+}\Gamma(1-q/\alpha_+)\Gamma(1+q/\alpha_+)}{ \cos(q\pi/2)\Gamma(1-q)\Gamma(1+q\alpha_-/\alpha_+)},  \\[4pt]
    M^{\prec}_2&=\int_{-\infty}^\infty|z|^q g_{\textrm{cen2}}(z)dz  \\
    & =\frac{(4D)^{q/2}\Gamma(\frac{q+1}{2})}{\sqrt{\pi}}, \\[4pt]
    M^{\succ}&=\int_{-\infty}^\infty|z|^q g_{\textrm{tail}}(z)dz  \\
    &=\frac{a_+q\Gamma(q-\alpha_+)}{2a_-\Gamma(q-\alpha_++\alpha_-+1)|\Gamma(1-\alpha_+)|}.
  \end{split}
\end{equation}
The coefficients $M^{\prec}_2$ and $M^{\succ}$ can be directly obtained by using the expressions of $g_{\textrm{cen2}}(z)$ and $g_{\textrm{tail}}(z)$, respectively. 
While it is not easy to get $M^{\prec}_1$ from the expression of $g_{\textrm{cen1}}(z)$, a Fox H-function. 
So we resort to the method of subordination and present the details in Appendix \ref{App2}.

\section{Summary}\label{eight}
The intermittent search strategy has been widely applied in the real world. The most powerful and representative one is an alternating process with two states: L\'{e}vy walk and Brownian motion. In this paper, we mainly investigate the anomalous diffusion with multiple modes for the two-state process. It is well-known that pure L\'{e}vy walk exhibits the strong anomalous diffusion when the power law exponent of running time is bigger than one. The intrinsic mechanism is that two kinds of distributions are complementary in the PDF of L\'{e}vy walk, i.e., the L\'{e}vy distribution in the central part and the infinite density in the tail part. Here, the two-state process becomes more complicated since three kinds of scales coexist in this system. The usual method to deal with the system with multiple modes is scaling analysis on the PDF. If $\alpha_->\alpha_+$ or $\alpha_->1$, the L\'{e}vy walk phase will dominate for long times in this system, and thus the strong anomalous diffusion phenomenon will be the same as pure L\'{e}vy walk. Therefore, we only consider the case $\alpha_-<\min(\alpha_+,1)$ in this paper.

Based on the technique of master equation, we build the integral equations for this two-state process, and thus obtain the explicit expression of PDF in Fourier-Laplace space $p(k,s)$ by solving the integral equations.  Consistent to the intuitive understanding of this system, three kinds of scaling regimes can be found in the expression of $p(k,s)$, which are $s\sim k$ for ballistic scaling in L\'{e}vy walk phase, $s\sim |k|^{\alpha_+/\alpha_-}$ for L\'{e}vy scaling in L\'{e}vy walk phase, and $s\sim |k|^2$ for Gaussian scaling in Brownian phase. By applying the detailed scaling analyses within these regimes, respectively, we obtain the infinite density in the tail part, and a combination of stretched L\'{e}vy and Gaussian distributions in the central part.

The relationships between the three distributions are abundant.
(i) In the central part, the leading role (with respect to moments) of stretched L\'{e}vy distribution and Gaussian distribution is determined by the magnitude size of $\alpha_+$ and $2\alpha_-$. The former distribution dominates when $\alpha_+<2\alpha_-$, otherwise the latter one dominates.
(ii) Whatever the magnitude sizes of $\alpha_+$ and $2\alpha_-$ are, it is the stretched L\'{e}vy distribution rather than Gaussian distribution, which is consistent to the infinite density in the intermediate region, since the Gaussian distribution decays exponentially and can be omitted.
(iii) There is a seeming accumulation effect at the end of the infinite density ($z=1$). The end of the infinite density is contributed by the particles running in its first step for the whole time. Once it renews and turns into the second step in Brownian phase, it is less likely for the particle to return the L\'{e}vy walk phase again due to the longer sojourn time in Brownian phase. Therefore, the PDF of L\'{e}vy walk is dropped down by the long sojourn time in Brownian phase except for the end point at $z=1$.
(iiii) Three distributions are equipped with different weights for different value of $q$ when calculating the absolute $q$-th moments. With their cooperation, all moments of the displacement $x(t)$ is finite and the strong anomalous diffusion can be observed.

\section*{Acknowledgments}
This work was supported by the National Natural Science Foundation of China under grant no. 11671182, and the Fundamental Research Funds for the Central Universities under grants no. lzujbky-2018-ot03 and no. lzujbky-2019-it17.

\appendix
\section{Inverse Fourier-Laplace transform of $R_\alpha(k,s)$ in Eq. \eqref{C1-Rks}}\label{App1}
First, the inverse Laplace transform ($s\rightarrow t$) of $R_\alpha(k,s)$ is
\begin{equation}
  \begin{split}
    R_\alpha(k,t)= -ikv_0 \int_0^t \frac{e^{-ikv_0t'}t'^{-\alpha_+}}{\Gamma(\alpha_-+1)\Gamma(1-\alpha_+)}(t-t')^{\alpha_-}dt'.
  \end{split}
\end{equation}
By performing the substitutions ($t'=tx$) and ($v_0tk=\xi$), one arrives at
\begin{equation}
  \tilde{R}_\alpha(\xi,t)= \frac{-i\xi t^{\alpha_--\alpha_+}}{\Gamma(\alpha_-+1)\Gamma(1-\alpha_+)}\int_0^1 e^{-i\xi x}x^{-\alpha_+}(1-x)^{\alpha_-}dx.
\end{equation}
Then, taking inverse Fourier transform ($\xi\rightarrow z$) leads to
\begin{equation}
\begin{split}
 &\tilde{R}_\alpha(z,t)  \\
 &= -\frac{t^{\alpha_--\alpha_+}}{\Gamma(\alpha_-+1)\Gamma(1-\alpha_+)} \frac{\partial}{\partial z}\int_0^1 \delta(z-x)x^{-\alpha_+}(1-x)^{\alpha_-}dx \\
    &= -\frac{t^{\alpha_--\alpha_+}}{\Gamma(\alpha_-+1)\Gamma(1-\alpha_+)} \frac{\partial}{\partial z}\Big[ \textbf{1}_{(0<z\leq1)}z^{-\alpha_+}(1-z)^{\alpha_-} \Big] \\
    &= \frac{t^{\alpha_--\alpha_+}}{\Gamma(\alpha_-+1)\Gamma(1-\alpha_+)} \mathcal{I}(z),
\end{split}
\end{equation}
where
\begin{equation}
  \mathcal{I}(z) = \textbf{1}_{(0<z\leq1)} z^{-\alpha_+-1}(1-z)^{\alpha_--1}\left[\alpha_++(\alpha_--\alpha_+)z \right].
\end{equation}
Finally, considering the relationship $v_0tk=\xi$, the inverse Fourier transform ($k\rightarrow x$) of $R_\alpha(k,t)$ is
\begin{equation}
\begin{split}
    R_\alpha(x,t) &=\frac{1}{v_0t} \tilde{R}_\alpha\left(\frac{x}{v_0t},t\right) \\
    &=\frac{t^{\alpha_--\alpha_+-1}}{v_0\Gamma(\alpha_-+1)\Gamma(1-\alpha_+)} \mathcal{I}\left(\frac{x}{v_0t}\right).
\end{split}
\end{equation}
Similarly, the inverse Fourier-Laplace transform ($k\rightarrow x,s\rightarrow t$) of $R_\alpha(-k,s)$ is
\begin{equation}
\begin{split}
     \frac{t^{\alpha_--\alpha_+-1}}{v_0\Gamma(\alpha_-+1)\Gamma(1-\alpha_+)} \mathcal{I}\left(-\frac{x}{v_0t}\right).
\end{split}
\end{equation}

\section{Derivation of the coefficient $M^{\prec}_1$}\label{App2}
It is not easy to directly obtain $M^{\prec}_1$ in Eq. \eqref{M-coefficients} from the expression of $g_{\textrm{cen1}}(z)$, since $g_{\textrm{cen1}}$ is a Fox H-function. However, we find that the PDF $p(x,t)$ in Eq. \eqref{C1-pks5}  corresponds to the model---L\'{e}vy flight coupled with an inverse subordinator, the Langevin picture of which is discussed in Ref. \cite{Fogedby:1994}. Based on the method of subordination \cite{BauleFriedrich:2005,ChenWangDeng:2018-2,ChenWangDeng:2019-2}, $p(x,t)$ can be written into an integral form as
\begin{equation}\label{B1}
  p(x,t)=\int_0^\infty p_0(x,\tau)h(\tau,t)d\tau,
\end{equation}
where $p_0(x,\tau)$ is the PDF of displacement of L\'{e}vy flight with Fourier transform ($x\rightarrow k$) being
\begin{equation}\label{B2}
  p_0(k,\tau)=e^{-\tau K_\alpha|k|^{\alpha_+}},
\end{equation}
and $h(\tau,t)$ is the PDF of the inverse $\alpha_-$-stable subordinator with Laplace transform ($t\rightarrow s$) being
\begin{equation}\label{B3}
  h(\tau,s)=s^{\alpha_--1}e^{-\tau s^{\alpha_-}}.
\end{equation}
The Eq. \eqref{C1-pks5} can be obtained by substituting the Eqs. \eqref{B2} and \eqref{B3} into Eq. \eqref{B1}. Multiplying $|x|^q$ on both sides of Eq. \eqref{B1}, we obtain
\begin{equation}\label{B4}
  \langle |x(t)|^q\rangle=\int_0^\infty \langle |x|_0^q\rangle(\tau)h(\tau,t)d\tau,
\end{equation}
where
\begin{equation}\label{B5}
  \langle |x(\tau)|_0^q\rangle =K_1 \tau^{q/\alpha_+}
\end{equation}
is the absolute $q$-th moment of L\'{e}vy flight. Here, \cite{SamorodnitskyTaqqu:1994,RebenshtokDenisovHanggiBarkai:2014-2}
\begin{equation}\label{B6}
\begin{split}
    K_1&=(K_\alpha)^{q/\alpha_+} \int_{-\infty}^\infty |x|^q l_{\alpha_+,0}(x)dx  \\
    &= \frac{(K_\alpha)^{q/\alpha_+}\Gamma(1-q/\alpha_+)}{\cos(q\pi/2)\Gamma(1-q)}
\end{split}
\end{equation}
with $l_{\alpha_+,0}(x)$ being the symmetric $\alpha_+$-stable L\'{e}vy noise \cite{KlafterSokolov:2011}. Substituting Eq. \eqref{B5} into Eq. \eqref{B4} and performing Laplace transform ($t\rightarrow s$), we obtain
\begin{equation}
  \mathcal{L}\{\langle |x(t)|^q\rangle\}=K_1 \Gamma(1+q/\alpha_+)\, s^{-1-q\alpha_-/\alpha_+},
\end{equation}
the inverse Laplace transform of which is
\begin{equation}
  \langle |x(t)|^q\rangle= M^{\prec}_1\,t^{q\alpha_-/\alpha_+}
\end{equation}
with
\begin{equation}
  M^{\prec}_1  =\frac{K_1 \Gamma(1+q/\alpha_+)}{\Gamma(1+q\alpha_-/\alpha_+)}.
\end{equation}

\section*{References}
\bibliographystyle{apsrev4-1}
\bibliography{ReferenceW}

\end{document}